\documentclass[conference]{IEEEtran}

\IEEEoverridecommandlockouts
\usepackage{verbatim}
\usepackage{longtable}
\DeclareMathAlphabet{\mathpzc}{OT1}{pzc}{m}{it}
\usepackage{amsmath}
\usepackage{amsfonts}
\usepackage{amssymb}
\usepackage{multirow}
\usepackage{amsmath,amssymb,amsfonts}
\usepackage{fancyhdr}
\usepackage[ruled,vlined,linesnumbered]{algorithm2e}
\usepackage[
top    = 0.7in,
bottom = 0.99in,
left   = 0.65in,
right  = 0.59in]{geometry}
\usepackage{graphicx}
\usepackage{textcomp}
\usepackage{xcolor}
\usepackage{subcaption}
\usepackage{url}

\usepackage{algpseudocode}
\usepackage{enumitem}
\usepackage{float}
\restylefloat{figure}
\usepackage{graphicx}
\usepackage[normalem]{ulem}
\useunder{\uline}{\ul}{}

\begin{document}

\title{Securing Proof of Stake Blockchains: Leveraging Multi-Agent Reinforcement Learning for Detecting and Mitigating Malicious Nodes}


\author{\IEEEauthorblockN{Faisal Haque Bappy$^{1}$, Tariqul Islam$^{2}$, Kamrul Hasan$^{3}$, Md Sajidul Islam Sajid$^{4}$, and Mir Mehedi Ahsan Pritom$^{5}$}
\IEEEauthorblockA{
$^{1, 2}$ Syracuse University, Syracuse, NY, USA\\
$ ^{3}$ Tennessee State University, Nashville, TN, USA\\
$ ^{4}$ Computer and Information Sciences, Towson University, Towson, MD, USA\\
$ ^{5}$ Computer Science, Tennessee Tech University, Cookeville, TN, USA\\
Email: \{fbappy@syr, mtislam@syr, mhasan1@tnstate, msajid@towson, mpritom@tntech\}.edu} 
}

\maketitle

\thispagestyle{fancy}
 \lhead{This work has been accepted at the 2024 IEEE Global Communications Conference (Globecom 2024)}
\cfoot{}

\begin{abstract}
Proof of Stake (PoS) blockchains offer promising alternatives to traditional Proof of Work (PoW) systems, providing scalability and energy efficiency. However, blockchains operate in a decentralized manner and the network is composed of diverse users. This openness creates the potential for malicious nodes to disrupt the network in various ways. Therefore, it is crucial to embed a mechanism within the blockchain network to constantly monitor, identify, and eliminate these malicious nodes without involving any central authority. In this paper, we propose MRL-PoS+, a novel consensus algorithm to enhance the security of PoS blockchains by leveraging Multi-agent Reinforcement Learning (MRL) techniques. Our proposed consensus algorithm introduces a penalty-reward scheme for detecting and eliminating malicious nodes. This approach involves the detection of behaviors that can lead to potential attacks in a blockchain network and hence penalizes the malicious nodes, restricting them from performing certain actions. Our developed Proof of Concept demonstrates effectiveness in eliminating malicious nodes for six types of major attacks. Experimental results demonstrate that MRL-PoS+ significantly improves the attack resilience of PoS blockchains compared to the traditional schemes without incurring additional computation overhead.
\end{abstract}


\section{Introduction}
Proof of Stake blockchains have garnered considerable attention as viable alternatives to traditional Proof of Work systems due to their scalability and energy efficiency advantages. Ethereum's recent shift to PoS marks a significant milestone in the evolution of blockchain technology. PoS allows Ethereum validators to secure the network and validate transactions based on the amount of cryptocurrency they hold, rather than relying on computational power \cite{asif2023shaping}. In PoS blockchains, consensus mechanisms rely on participants' stakes in the network rather than computational power, offering a more environmentally sustainable approach to blockchain validation. However, the decentralized nature of PoS blockchains introduces inherent challenges, particularly in terms of security and resilience against malicious actors \cite{d2022no, mancino2023exploiting, pavloff2023ethereum}. As diverse user nodes interact within the network, the potential for malicious behaviors to disrupt operations poses a significant threat to the integrity and stability of PoS-based systems.

The openness and inclusivity of PoS blockchains, while beneficial for democratizing access and participation, also create opportunities for exploitation by malicious nodes \cite{pavloff2023ethereum}. These nodes may seek to undermine the network through various means, including double spending attacks, Sybil attacks, and DDoS attacks \cite{schwarz2022three}. The absence of centralized authority in PoS blockchains necessitates the implementation of robust mechanisms to detect, deter, and mitigate the impact of such malicious activities autonomously.

In response to these challenges, this paper proposes MRL-PoS+, a novel consensus algorithm that leverages Multi-agent Reinforcement Learning techniques to enhance the security of PoS blockchains. By integrating MRL principles into the consensus process, MRL-PoS+ aims to empower nodes within the network to actively identify and respond to malicious behaviors without centralized oversight. Through a penalty-reward framework, MRL-PoS+ incentivizes honest participation while penalizing and restricting the actions of malicious nodes, thereby fortifying the resilience of PoS blockchains against potential attacks. Through empirical analysis and experimentation, we demonstrate the efficacy of MRL-PoS+ in enhancing the security and robustness of PoS-based blockchains. The following are the core contributions of this paper,

\begin{itemize}
    \item We introduced MRL-PoS+, a novel consensus algorithm leveraging Multi-agent Reinforcement Learning techniques to secure the blockchain network without having any central authority.
    \item We implemented a penalty-reward mechanism within MRL-PoS+ to incentivize honest behaviors and penalize malicious nodes.
    \item We validated MRL-PoS+ through a Proof of Concept (PoC) implementation, demonstrating effectiveness against six major types of attacks.
    \item Empirical analysis highlights that MRL-PoS+ performs better than traditional schemes in enhancing network security without adding computational overhead.
\end{itemize}

The remainder of the paper is structured as follows: in Section \ref{sec:relatedworks}, we present existing works related to the security of PoS blockchain and reinforcement learning. In Section \ref{sec:attacks}, we described six major attacks in blockchain and their potential indicators. In Section \ref{sec:rl}, we explained how MRL-PoS+ integrates reinforcement learning inside a PoS blockchain network. Then in Section \ref{sec:system}, we present our system architecture, followed by a performance analysis in Section \ref{sec:performance}. Finally, Section \ref{sec:conclusion} concludes the paper.

\section{Related Works}
\label{sec:relatedworks}
In recent years, researchers have been exploring the integration of blockchain technology with multi-agent reinforcement learning to enhance the security, efficiency, and reliability of decentralized systems. Several studies \cite{zou2023optimized, li2023q, de2020blockchain, cachin2017blockchain}, have investigated consensus mechanisms and security measures in blockchain networks. Zou et al. \cite{zou2023optimized} propose a consensus protocol leveraging reinforcement learning to optimize efficiency and fairness, particularly in IoT networks. Li et al. \cite{li2023q} introduce a framework for trust management in IoT using blockchain, emphasizing credibility improvement and energy conservation. Oliveira et al. \cite{de2020blockchain} explore a reputation-based consensus mechanism to enhance security against adversarial attacks. 

Meanwhile, Cachin et al. \cite{cachin2017blockchain} review various consensus protocols across blockchain platforms, emphasizing fault tolerance and resistance against malicious actors. Several other studies explore the integration of blockchain and multi-agent systems, often with a focus on reinforcement learning. Sami et al. \cite{sami2024learnchain} propose utilizing blockchain for transparent and auditable model updates in multi-agent reinforcement learning. Calvaresi et al. \cite{calvaresi2018multi} advocate for combining blockchain and multi-agent systems to enhance trust and accountability. Nguyen et al. \cite{nguyen2021cooperative} propose a cooperative task offloading scheme for blockchain-based mobile edge computing, leveraging reinforcement learning for resource allocation. Shen et al. \cite{shen2023blockchain} develop a distributed multi-agent reinforcement learning-based algorithm for multi-object tracking, integrated into a blockchain framework. While these studies demonstrate the potential synergy between blockchain and multi-agent reinforcement learning, none explicitly addresses the utilization of MRL for malicious node detection and defense within blockchain networks.


\section{Attacks in Blockchain}
\label{sec:attacks}
In this section, we present an overview of 6 prominent blockchain attacks, their typical symptoms as documented in existing literature, and the detection criteria that contribute to the detection mechanisms employed in MRL-PoS+.

\subsection{51\% Attack}
The 51\% Attack poses a significant threat to blockchain integrity. It occurs when a single entity or coalition controls over half of the network's computational power, allowing them to manipulate transaction history and potentially disrupt consensus mechanisms \cite{aponte202151}. Symptoms include sudden fluctuations in network hash rate, frequent forks or reorganizations of recent blocks, and conflicting transactions within the blockchain \cite{sayeed2019assessing}. 

\textbf{Detection Criteria.} i) Analyzing block generation against a predefined threshold, ii) comparing node hash rates to the network average, and iii) calculating fork frequency collectively determine the likelihood of a 51\% attack.

\subsection{Double-Spending Attack}
The Double-Spending Attack remains a persistent concern for blockchain networks, exploiting transaction confirmation delays to spend the same cryptocurrency units multiple times. This attack can exploit consensus vulnerabilities or manipulate transaction records to achieve its objectives \cite{karpinski2021blockchain}. Symptoms of a Double-Spending Attack manifest as the presence of identical transaction IDs in multiple unconfirmed blocks, sudden discrepancies in account balances, and anomalous transaction histories \cite{zhang2019double}. 

\textbf{Detection Criteria.} i) Analyzing the number of duplicate transactions, ii) evaluating the history of double spending from unconfirmed blocks, and iii) detecting discrepancies in balance.

\subsection{Sybil Attack}
The Sybil Attack involves the creation of numerous fake identities or nodes to exert control or influence over a network. Attackers leverage this tactic to disrupt consensus, manipulate network voting processes, or inundate the network with spurious data \cite{saad2020exploring}. Symptoms of a Sybil Attack include a sudden surge of network participants with similar IP addresses or behavioral patterns, as well as an influx of irrelevant data overwhelming the network \cite{zhang2019double}.

\textbf{Detection Criteria.} i) Monitoring for abrupt spikes in the network activity, ii) monitoring the presence of spam, and iii) scrutinizing whether nodes are attempting to manipulate voting data.

\subsection{Replay Attack}
A Replay Attack occurs when a legitimate data transmission is maliciously repeated or delayed by an adversary. In the blockchain context, attackers exploit vulnerabilities to resubmit valid transactions multiple times, potentially leading to unauthorized actions or financial losses \cite{duan2022multiple}. Symptoms of a Replay Attack include the appearance of duplicate transactions with differing timestamps and unexplained spikes in transaction fees \cite{zhang2019double}.

\textbf{Detection Criteria.} i) Scrutinizing for duplicate transactions within the network, ii) detecting a large number of conflicting transactions, and ii) considering any sudden or abnormal increases in transaction fees, indicating the presence of a probable replay attack.

\subsection{Smart Contract Vulnerabilities}
Smart contracts, while offering automated and trustless execution of agreements, are susceptible to exploitation due to coding errors or vulnerabilities. Malicious actors may exploit these weaknesses to extract funds, disrupt operations, or manipulate contract outcomes \cite{sayeed2020smart}. Symptoms of Smart Contract exploitation encompass irregular movements of funds, abnormal activity surrounding specific contracts, and unexpected alterations in contract behavior \cite{duan2022multiple}. 

\textbf{Detection Criteria.} i) Detecting any large unexplained transfers, and ii) identifying excessive activities on a single contract.

\subsection{DDoS Attack}
Distributed Denial of Service (DDoS) attacks can disrupt network operations, impede transaction processing, and degrade overall system performance \cite{mirkin2020bdos}. Symptoms of a DDoS Attack include frequent network errors, connectivity failures, and resource depletion on affected nodes \cite{mirkin2020bdos}.

\textbf{Detection Criteria.} i) Examining excessive network errors and overuse of resources, and ii) identifying anomalies suggesting a node's attempt to disrupt transaction processing.

\section{Reinforcement Learning inside a Blockchain Network}
\label{sec:rl}
In PoS Blockchains, nodes compete to process transactions by staking assets, with the selection of a validator determined pseudo-randomly based on factors like stake amount and transaction history. In our proposed MRL-PoS+ approach, we introduce a reinforcement learning process for blockchain nodes. Since these nodes are operated by individual users seeking incentives, complete automation of their actions is not viable. Thus, we view them as semi-automated agents for reinforcement learning instead of fully automated ones. 

\subsection{Multi-Agent Reinforcement Learning}
In multi-agent reinforcement learning (MRL), each agent's goal is typically to maximize its own cumulative reward, but the agents must learn to balance their individual goals with the overall goals of the group or system. The basic equation used in our proposed MRL-PoS+, known as the Q-learning algorithm, guides how agents update their strategies over time \cite{yang2020overview}. 
\begin{equation}
    Q(s, a) \leftarrow (1 - \alpha) \cdot Q(s, a) + \alpha \cdot (r + \gamma \cdot max_{a'} Q(s', a'))
\end{equation}
Where, $Q(s, a)$ is the Q-value, $\alpha$ is the learning rate, $r$ is the reward, $\gamma$ is the discount factor, $s'$ is the next state, and $\max_{a'} Q(s', a')$ is the maximum Q-value for the next state.


\subsection{Semi-Automated Agents} 
In traditional reinforcement learning, agents autonomously learn through network interaction. In our proposal, blockchain nodes are viewed as semi-automated agents, allowing human operators some control. Each node manages a reputation table, treating its transaction-related activities as actions. Nodes receive penalties or rewards, like allowances or action restrictions, based on their behavior, incentivizing honesty to earn rewards. Failure to comply may lead to exclusion from consensus, rendering the node ineffective.

\subsection{Set of Actions} 
Considering the entire blockchain network as the environment, a validator's main tasks, like transaction processing and block validation, form the set of actions. However, certain behaviors during these tasks may indicate malicious intent or an attack attempt. To bolster PoS blockchain security, such actions are subject to penalties. Building on the outlined attacks, we've identified 16 behaviors as malicious, listed comprehensively in Table \ref{tab:Malicious Behaviors}.

\begin{table}[]
\centering
\caption{List of Malicious Behaviors}
\label{tab:Malicious Behaviors}
\resizebox{\columnwidth}{!}{%
\begin{tabular}{cl}
\hline
\textbf{Identifiers}  & \multicolumn{1}{c}{\textbf{Behaviors}}                                                                                         \\ \hline
\textit{\textbf{b1}}  & \begin{tabular}[c]{@{}l@{}}One node consistently generates a majority\\ of new blocks\end{tabular}                             \\ \hline
\textit{\textbf{b2}}  & \begin{tabular}[c]{@{}l@{}}Sharply increasing or decreasing hash rate, \\ potentially exceeding network capacity\end{tabular}  \\ \hline
\textit{\textbf{b3}}  & \begin{tabular}[c]{@{}l@{}}Frequent forks or rewrites of recent blocks, \\ possibly with conflicting transactions\end{tabular} \\ \hline
\textit{\textbf{b4}}  & \begin{tabular}[c]{@{}l@{}}Same transaction ID appearing in multiple \\ unconfirmed blocks\end{tabular}                        \\ \hline
\textit{\textbf{b5}}  & \begin{tabular}[c]{@{}l@{}}Two transactions with the same coins spent \\ at different times\end{tabular}                       \\ \hline
\textit{\textbf{b6}}  & \begin{tabular}[c]{@{}l@{}}Sudden drops or discrepancies in account \\ balances, particularly after confirmation delays.\end{tabular}       \\ \hline
\textit{\textbf{b7}}  & \begin{tabular}[c]{@{}l@{}}A sudden increase in network participants, \\ often with similar IP addresses or voting patterns.\end{tabular}   \\ \hline
\textit{\textbf{b8}}  & Manipulation of voting power                                                                                                   \\ \hline
\textit{\textbf{b9}}  & \begin{tabular}[c]{@{}l@{}}Spam or flooding of the network with irrelevant \\ data\end{tabular}                                \\ \hline
\textit{\textbf{b10}} & \begin{tabular}[c]{@{}l@{}}Same transaction appearing multiple times in the \\ blockchain, often with different timestamps\end{tabular}     \\ \hline
\textit{\textbf{b11}} & Conflicting transaction confirmations                                                                                          \\ \hline
\textit{\textbf{b12}} & \begin{tabular}[c]{@{}l@{}}Sudden increases in transaction fees without any \\ apparent reason\end{tabular}                    \\ \hline
\textit{\textbf{b13}} & Large and unexplained fund movements                                                                                           \\ \hline
\textit{\textbf{b14}} & \begin{tabular}[c]{@{}l@{}}Abnormal activity around specific smart contracts, \\ potentially indicating exploitation attempts.\end{tabular} \\ \hline
\textit{\textbf{b15}} & Frequent network errors/failures to connect to nodes                                                                           \\ \hline
\textit{\textbf{b16}} & Server overload and resource depletion on nodes                                                                                \\ \hline
\end{tabular}
}
\end{table}

\subsection{Penalizing Malicious Nodes and Rewarding Honest Nodes} 
In the event that a node exhibits any of the identified malicious behaviors during transaction processing, the system will impose penalties on the node. Additionally, nodes will be subjected to certain restrictions corresponding to their displayed behavior. To record the penalty, the node's reputation table will be updated, and the election process will take into account the reputation tables of all nodes. Nodes that do not exhibit any malicious behavior are deemed honest nodes and are eligible for rewards within the system. The magnitude of rewards varies depending on the node's state: i) if a node has never previously engaged in malicious behavior, it will receive the highest amount of reward, ii) for nodes with a history of previous malicious behavior, their negative reputations will be mitigated. Additionally, rewards will grant these nodes the ability to undertake certain actions that were previously restricted.

\subsection{Malicious Behavior Detection}
To integrate the penalty-reward mechanism into the blockchain network, the initial challenge is monitoring and tracking every node action during transaction processing. In MRL-PoS+, we introduced attack probability detection mechanisms for six major attack types. A node's total attack probability aggregates probabilities of all attacks ($b1, b2, b3, ..., b16$), weighted by predefined criteria. These weights may be predetermined by network policy or dynamically adjusted based on consensus among participating nodes. By considering these probabilities, the system can make informed decisions on node selection.

\section{System Architecture}
\label{sec:system}
\begin{figure}[]
  \centering
  \includegraphics[width=\columnwidth]{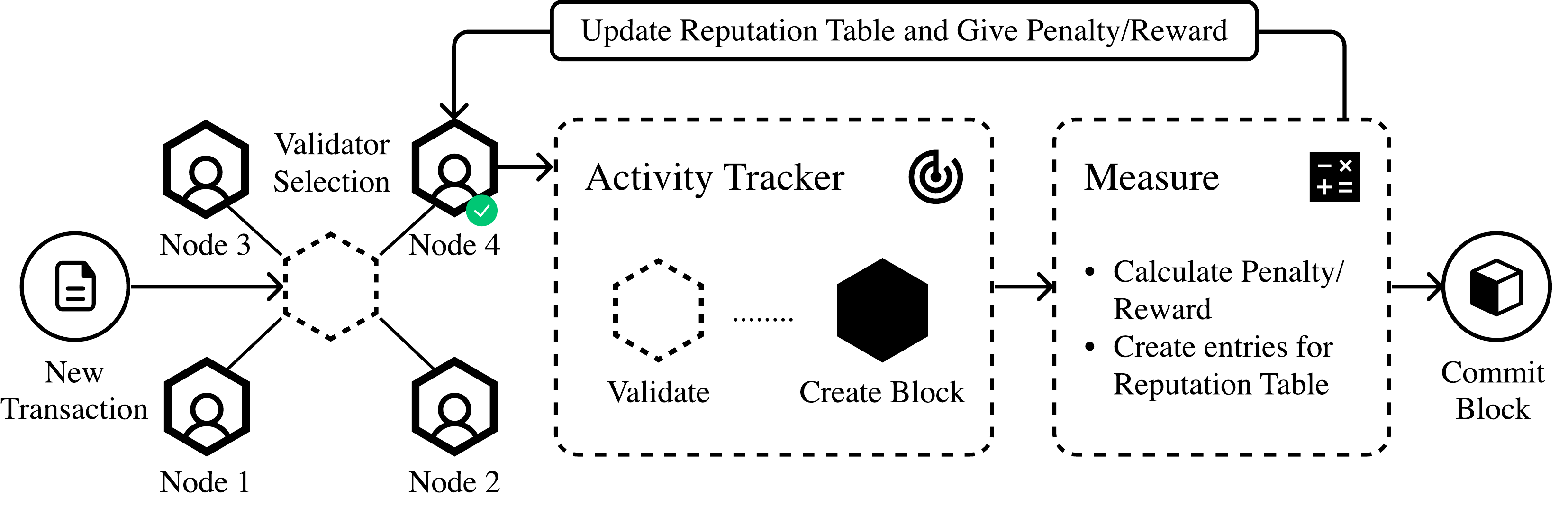}
  \caption{System Architecture of MRL-PoS+}
  \label{fig:sysArch}
\end{figure}

As we prioritize PoS blockchain security, MRL-PoS+ mirrors traditional PoS systems in architecture. Figure \ref{fig:sysArch} outlines MRL-PoS+'s workflow: when a user initiates a transaction, network nodes compete to validate it. Unlike PoS, MRL-PoS+ factors node reputation into selection alongside stake. Selected validators, like Node 4, process transactions to create blocks. MRL-PoS+ integrates an activity tracker to monitor and record node behaviors without additional data storage for decentralization. The node structure is updated to accommodate tracked parameters. After tracking, MRL-PoS+ calculates penalties and rewards based on node actions, updates reputation tables, and commits blocks to the main chain. This comprehensive approach ensures blockchain integrity and security, encouraging honest node participation.



\subsection{Activity Tracker}
The activity tracker detects malicious activities within the network, employing various detection mechanisms. It operates efficiently, eliminating the need for continuous monitoring by executing relevant functions after each action in the block creation process. This streamlined approach ensures swift detection and response to any suspicious behavior, bolstering the security and integrity of the blockchain network.

\subsection{Reputation Table}
Each node within the network possesses a reputation table structured as key-value pairs. Upon joining the network, a new node initializes all parameters with empty or zero values. The reputation table comprises four key parameters: i) $attackProbability$: aggregates probabilities from $b1$ to $b16$ (refer to Table \ref{tab:Malicious Behaviors}); ii) $lastAttackAge$: assigned based on a predetermined learning rate, influencing the node's ability to recover reputation swiftly by exhibiting honest behavior; iii) $restrictions$: a list of prohibited actions serving as penalties for malicious behaviors, with specific restrictions tied to exhibited malicious behaviors; iv) $isActive$: a binary parameter indicating the node's active participation in the consensus. If a node consistently engages in malicious activities, causing a substantial reputation loss, it becomes ineligible for the validator selection process and is marked inactive. Figure \ref{fig:q-learn} depicts a node's reputation table evolution over multiple rounds, reflecting parameter changes over time influenced by the node's actions and network dynamics.

\begin{figure}[]
  \centering
  \includegraphics[width=0.9\columnwidth]{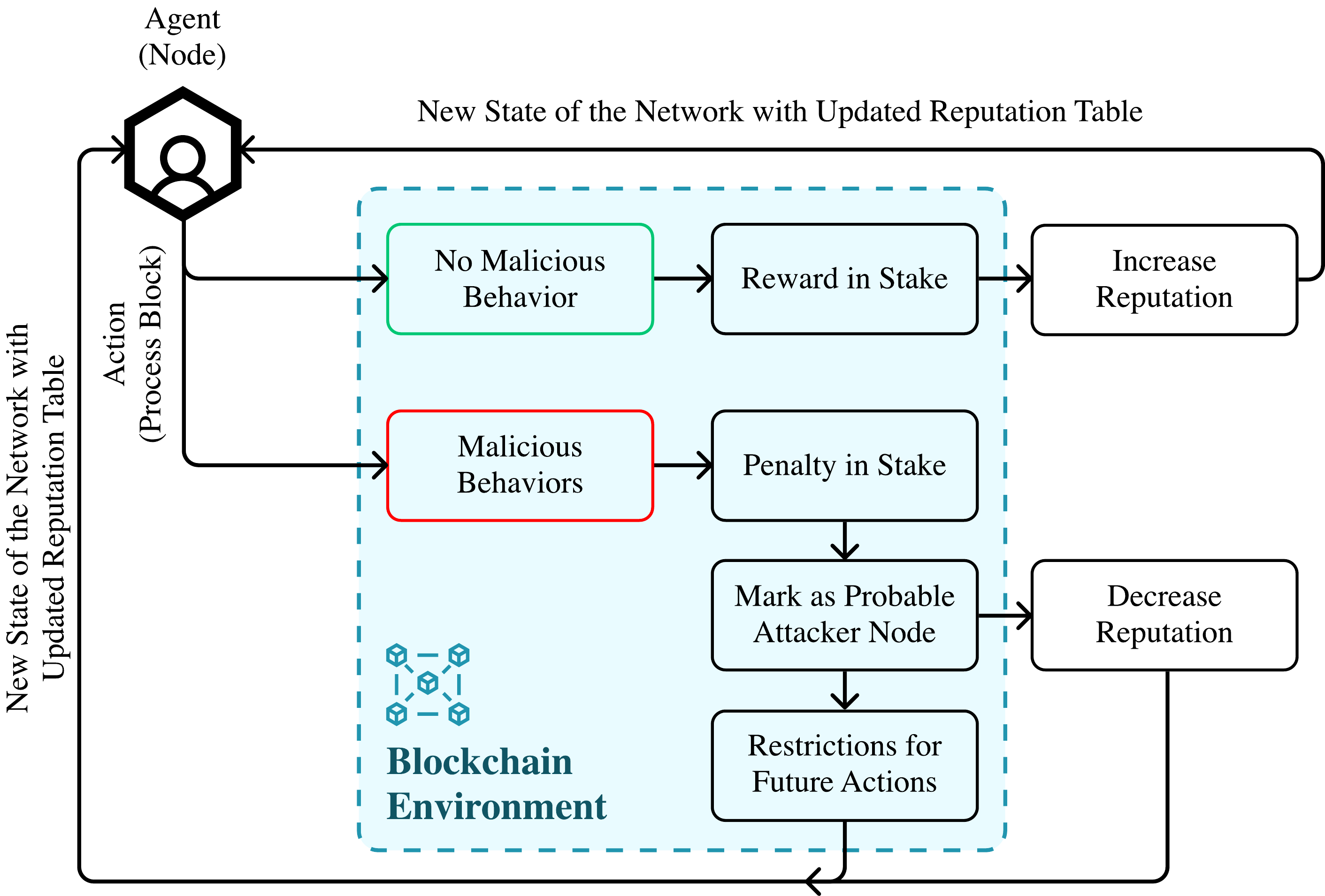}
  \caption{Penalty Reward Mechanism}
  \label{fig:penaltyReward}
\end{figure}

\begin{figure*}[h]
\centering
\includegraphics[width=\linewidth]{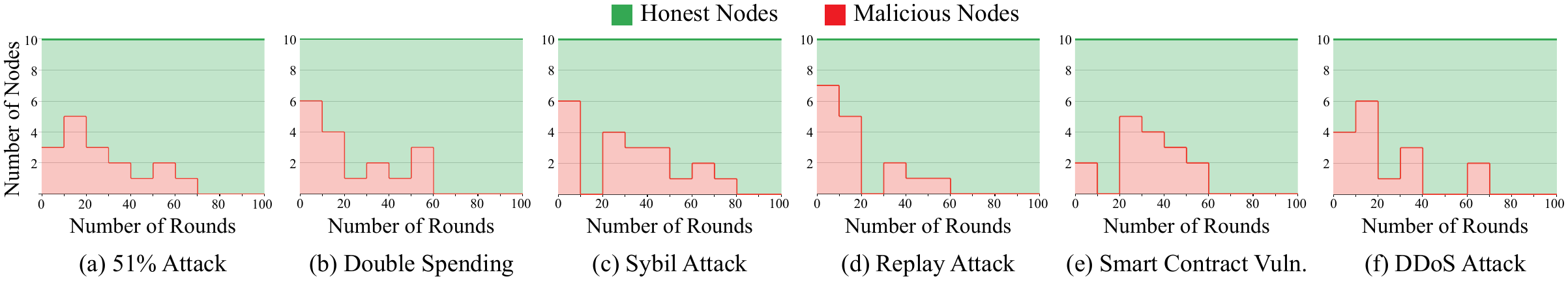}
\caption{Elimination of Malicious Nodes for Different Attacks}
\label{fig:attack}
\end{figure*}

\subsection{Validator Selection}
In MRL-PoS+, when nodes express interest in processing a transaction, they commit their stake and engage in a selection process. The number of participants may vary depending on the incentive associated with each transaction. Algorithm \ref{alg:elect} outlines the validator selection process within MRL-PoS+. For each node, it first accesses the total attack probability from the reputation table. Then, it considers the age of the last attack and determines if the product of the attack's age and the attack probability exceeds a certain threshold value. If so, the node is excluded from selection, though its attack age decreases as it seeks consensus participation. For the remaining nodes, their agent score is calculated by considering attack probability, attack age, and stake. Based on this score, the nodes are sorted. If multiple nodes have the same highest stake, one of them will be chosen randomly. Otherwise, the node with the highest agent score will be chosen as the validator.

\IncMargin{1em}
\setlength{\textfloatsep}{0pt}
 \begin{algorithm}
  \caption{ElectValidator()}\label{alg:elect}
  \SetKwFunction{BuildTree}{ElectValidator}
  \SetKwInOut{Input}{Input}
  \SetKwInOut{Output}{Output}
  \Indm 
    \Input{Participating Nodes $nodes$} 
    \Output{Elected Node $N$}
    \Indp
    \BlankLine
    \ForEach{$n$ in $nodes$}{
        $atkPr \leftarrow \texttt{getAtkProb}(n)$\\
        $lastAtk \leftarrow n.atkAge$ \Comment{$atkAge$ can be $0, 1, 2, ..., n$ depending on the learning rate. It means the node attempted an attack $(n-atkAge)$ rounds ago}\\
        \If{$atkProb * lastAtk < threshold$}{
            $n.score \leftarrow -1 * atkProb * lastAtk * n.stake$\\
            $electedAgents.\texttt{push}(n)$\\
        }
        \If{$n.atkAge >0$}{
            $n.atkAge -= 1$\\
        }
    }
    $electedAgents.\texttt{sort}(by=score)$\\
    \If{$\texttt{hasMultipleHighestStake}()$}{
        return $\texttt{randomNodeWithHighestStake}()$\\
    }
    \Else{
        return $electedAgents[0]$\\
    }
\end{algorithm}
\DecMargin{1em}

\subsection{Penalty-Reward Mechanism}
The process of assigning penalties or rewards is shown in Figure \ref{fig:penaltyReward}. Initially, the system checks whether the node has exhibited any malicious behaviors. If not, the node is considered honest and receives its incentives from the transaction fee. Additionally, the node's reputation values are updated: if it has any attack probability, it is decreased, and the values of $b1$ to $b16$ are set to negative. This prioritizes the node over less reputable nodes in future rounds. Contrarily, if the node engages in malicious behavior, its stake is retained as a penalty. Also, its reputation is diminished, with a higher attack probability assigned. Furthermore, restrictions are imposed on the node's actions, preventing it from attempting the same malicious behavior in subsequent transactions. 

\section{Performance Analysis}
\label{sec:performance}
In this section, we present the experimental results obtained with MRL-PoS+. We utilized our developed proof of concept to evaluate effectiveness and performance through several metrics.

\subsection{Simulation Environment}

To evaluate our proposed consensus mechanism's functionality and performance, we developed a custom blockchain network using Go, similar to Ethereum's setup\footnote{The implementation of our proposed MRL-PoS+ is available on: \\  \url{https://github.com/SPaDeS-Lab/mrl-pos-plus}}, ensuring compatibility with its PoS blockchain model, the current most popular. Our experiments deliberately varied node numbers to explore how our method operates under diverse network configurations, yielding insights into scalability, reliability, and efficiency. All tests were conducted on Azure VMs (4 vCPUs, 16GB memory) running on Ubuntu 22.04, each hosting 4 containerized nodes, enabling simulation of real-world conditions and thorough analyses in a controlled environment.

\subsection{Detection and Elimination of Malicious Nodes}
In our initial test, we evaluated MRL-PoS+'s ability to detect and remove malicious nodes using its penalty-reward system by simulating nodes engaging in various attacks, each specializing in a specific type. Figure 3 displays the results, showing the number of malicious (in red) and honest (in green) nodes selected as validators, aggregated every 10 rounds. Each simulation included 40 malicious and 60 honest nodes per attack type, maintaining consistent behavior over 100 rounds. This rigorous setup tests MRL-PoS+'s performance under challenging conditions, recognizing the difficulty of sustaining consistent malicious behavior due to nodes' limited stakes in real networks. The plots in Figure 3 consistently show a decline in malicious validators as rounds progress, attributed to honest nodes gaining reputation and rewards while malicious ones face penalties, resulting in stake loss, restrictions, and reputation decline. This trend persists across all attack types, showcasing MRL-PoS+'s effectiveness in detecting and removing malicious nodes in challenging scenarios, highlighting its relevance and resilience in securing PoS blockchains.

\subsection{Learning Stages}
\begin{figure}[]
  \centering
  \includegraphics[width=\columnwidth]{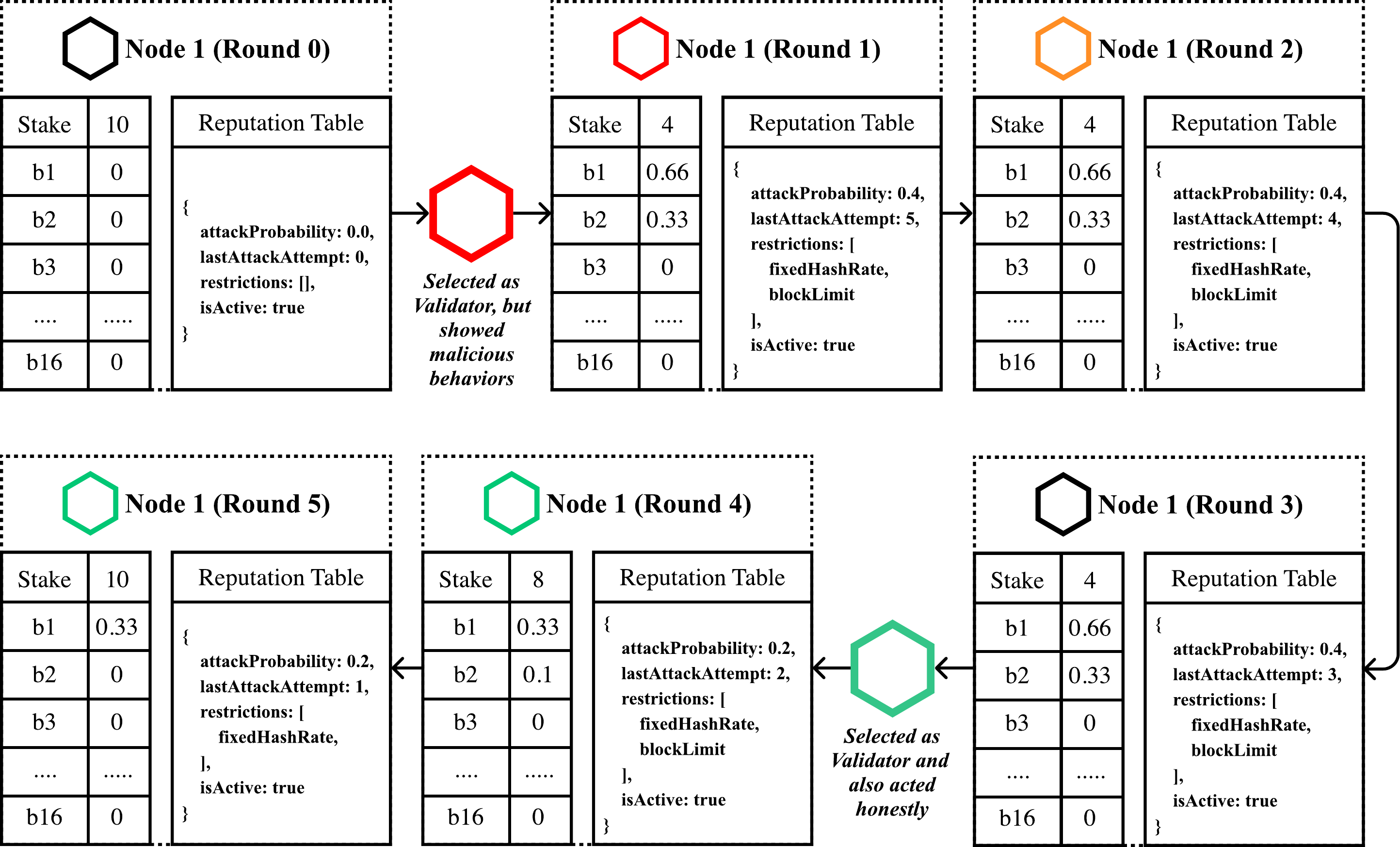}
  \caption{Learning Stages of a Regular Node}\label{fig:q-learn}
\end{figure}

The preceding experiment showcased an extreme scenario where all malicious nodes consistently behaved maliciously. However, in reality, nodes are expected to recover and return to honest participation. To illustrate this, we simulated a typical node scenario in MRL-PoS+. Figure \ref{fig:q-learn} illustrates the evolution of the reputation table and stake across six rounds.

\begin{figure}[h]
  \centering
  \includegraphics[width=0.85\columnwidth]{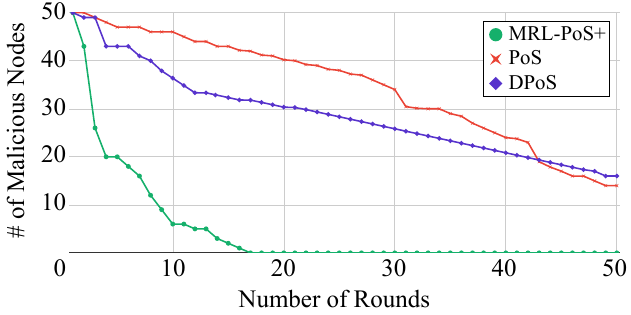}
  \caption{Comparison with Regular PoS and DPoS}\label{fig:comparison}
\end{figure}
In round 0, a node enters the network and participates in the initial validator selection, only to engage in malicious behavior upon selection, resulting in a penalty and its classification as an attacker node. This penalty also imposes limitations on the node's reputation table. Despite the penalty, the node persists in the selection process for three more rounds without success, with each round reducing the attack age. In round 3, the node is once again chosen as a validator and behaves honestly this time. Consequently, it regains some reputation in round 4 and sheds a restriction in the subsequent round. Continued honest behavior will lead to rewards and a positive reputation. However, any subsequent malicious actions will incur significant penalties, ultimately resulting in stake loss and exclusion from the selection process.

\subsection{Comparison with Traditional Approaches}
While traditional PoS and DPoS \cite{saad2020comparative} lack attack detection mechanisms, they seize the stake of nodes failing to process transactions correctly. In this experiment, we simulated PoS, DPoS, and MRL-PoS+ with 50 malicious nodes. We tracked how many rounds it took for these nodes to lose their stakes, considering a balance below 0 as elimination. Figure \ref{fig:comparison} demonstrates MRL-PoS+'s significantly quicker elimination of malicious nodes compared to PoS and DPoS, which fail to eradicate them all. This discrepancy arises because not all malicious behaviors cause system failures, thus PoS and DPoS do not detect them and withhold incentives. Consequently, after 50 rounds, more than 10 malicious nodes persist in PoS and DPoS networks. Although DPoS slightly outperforms PoS, it remains ineffective in eliminating malicious nodes. During our experiments, we measured the average memory usage. In both cases, the average usage was ~60\% (9.7GB for PoS and 9.81GB for MRL-PoS+). Although MRL-PoS+ is based on PoS and adds some extra layers to it, it does not increase the overhead.



\section{Conclusion}
\label{sec:conclusion}
In this paper, we introduce MRL-PoS+, a novel consensus algorithm designed to enhance the security of PoS blockchains. Leveraging Multi-agent Reinforcement Learning techniques, MRL-PoS+ incorporates a penalty-reward mechanism to incentivize honest behavior and penalize malicious nodes. By providing empirical evidence of its efficacy, this paper contributes to the ongoing discourse on blockchain security and consensus algorithm design. Furthermore, the persuasiveness of MRL-PoS+ makes it a promising solution for addressing the security challenges inherent in PoS blockchains while ensuring the integrity and resilience of decentralized networks. Moving forward, future research may explore additional enhancements and optimizations to further strengthen MRL-PoS+ and adapt it to evolving security threats in the blockchain landscape.

\bibliographystyle{IEEEtran}
\bibliography{IEEEabrv,references}

\end{document}